\documentclass[english,aps,prb,floats,twocolumn]{revtex4}
\usepackage[T1]{fontenc}
\usepackage[latin9]{inputenc}
\usepackage{bm}
\usepackage{amsmath}
\usepackage{amssymb}
\usepackage{graphicx}
\usepackage{esint}
\usepackage{adjustbox}
\usepackage{longtable}
\makeatletter
\@ifundefined{textcolor}{}
{%
 \definecolor{BLACK}{gray}{0}
 \definecolor{WHITE}{gray}{1}
 \definecolor{RED}{rgb}{1,0,0}
 \definecolor{GREEN}{rgb}{0,1,0}
 \definecolor{BLUE}{rgb}{0,0,1}
 \definecolor{CYAN}{cmyk}{1,0,0,0}
 \definecolor{MAGENTA}{cmyk}{0,1,0,0}
 \definecolor{YELLOW}{cmyk}{0,0,1,0}
}

\usepackage{babel}

\makeatother

\usepackage{babel}
\begin{document}

\title{Dynamics of Skyrmion Contraction and Expansion in a Magnetic Film}

\author{Eugene M. Chudnovsky}

\affiliation{Physics Department, Herbert H. Lehman College and Graduate School,
The City University of New York, 250 Bedford Park Boulevard West,
Bronx, New York 10468-1589, USA }

\date{\today}
\begin{abstract}
Contraction and expansion of skyrmions in ferromagnetic films are investigated. In centrosymmetric systems, the dynamics of a collapsing skyrmion is driven by dissipation. The collapse time has a minimum on the damping constant. In systems with broken inversion symmetry, the evolution of skyrmions toward equilibrium size is driven by the Dzyaloshinskii-Moriya interaction. Expressions describing the time dependence of the skyrmion size are derived and their implications for skyrmion-based information processing are discussed.  \\
\\
\footnotesize{Keywords: ferromagnetic films, skyrmion expansion and collapse, speed of skyrmion-based information processing}

\end{abstract}

\maketitle

\section{Introduction} \label{Intro}

Skyrmions found their way to material science from nuclear physics where they were introduced as models of nucleons \cite{SkyrmePRC58,Polyakov-book}. The similarity of the $\sigma$-model to the continuous spin-field model of the exchange interaction prompted studies of skyrmions in theories of ferro- and antiferromagnets \cite{BP,Manton-book,Lectures}. Topological arguments that produce skyrmions also arise in Bose-Einstein condensates \cite{AlkStoNat01}, quantum Hall effect \cite{SonKarKivPRB93,StonePRB93}, anomalous Hall effect \cite{YeKimPRL99}, liquid crystals \cite{WriMerRMR89} and graphene \cite{graphene}.

In recent years, studies of skyrmions in thin magnetic films, besides their fundamental value, have been inspired by the prospect of developing skyrmion-based topologically-protected data storage and information processing \cite{Leonov-NJP2016,Fert-Nature2017,Bogdanov2020,Luo2021}. Crucial to this task is the dynamics of the creation and annihilation of skyrmions \cite{Romming,Berruto-PRL2018,McGray, Garanin-JAP2018,Zhang-APL2018,Zhang2020,Muckel2021}. Characteristic times involved in such processes will determine the competitiveness of the skyrmion-based computer technology if it ever materializes. 

In a continuous field theory, the stability of a textbook Belavin-Polyakov skyrmion \cite{BP} against contraction or expansion is due to the scale invariance of the 2D exchange model. Atomic lattice breaks this invariance. In real magnetic films, skyrmions must be stabilized by interactions other than Heisenberg exchange. Most often it is the combined effect of the magnetic field and Dzyaloshinskii-Moriya (DMI) interaction \cite{Bogdanov1989,Bogdanov94,Bogdanov-Nature2006}. Other mechanisms of skyrmion stabilization include frustrated exchange \cite{Leonov-NatCom2015,Zhang-NatCom2017}, magnetic anisotropy \cite{Kovalev1979,IvanovPRB06,Lin-PRB2016}, disorder \cite{CG-NJP2018}, and geometrical confinement \cite{Moutafis-PRB2009}.

The stability of a skyrmion to thermal fluctuations capable of kicking it out of a metastable state has been studied by several authors \cite{bessarab18,stosic17,desplat18,hagemeister15,rohart16,rozsa16,siemens16,Malottki-PRB2019,Amel-JAP2019,Hoffmann-PRL2020}. Similar effect due to quantum fluctuations \cite{MQT-book} has been investigated as well \cite{Psaroudaki-PRX2017,Amel2018,AFM}. These works addressed the lifetime of a skyrmion in a metastable state but did not study its evolution in time after it goes over or under the energy barrier. 

The root problem of that kind is the collapse of a BP skyrmion due to effects unaccounted for in a continuous-field 2D exchange model, such as, e.g., the Landau-Lifshitz (LL) damping \cite{Abanov}. In addition,  the nonlinearity of the BP model on a lattice, and the emission of spin waves caused by it, leads to the effective dissipation of the skyrmion motion that emulates the LL damping \cite{DCG-PRB2022}. This problem will be reviewed here and the formulas describing skyrmion collapse in the case of an arbitrary strength of the LL damping will be derived. 

Then we will move to the dynamics of a nucleated skyrmion which allows it to expand toward the equilibrium size determined by the magnetic field in the model with DMI. Formulas describing the time evolution of such a skyrmion will be derived. Characteristic times required for a skyrmion to stabilize will be estimated. They are important for evaluating the prospect of building skyrmion-based information technology. We will argue that the use of high magnetic fields is crucial for that purpose.

The article is organized as follows. Skyrmion on a lattice is considered in Section \ref{lattice}. Dissipation is introduced in Section \ref{dissipation}. The model with DMI is studied in Section \ref{DMI}. Section \ref{conclusion} contains a brief summary of the results and comments on their applications. 

\section{Skyrmion on a lattice}
\label{lattice}

\begin{figure}[h]
\centering{}\includegraphics[width=8.5cm]{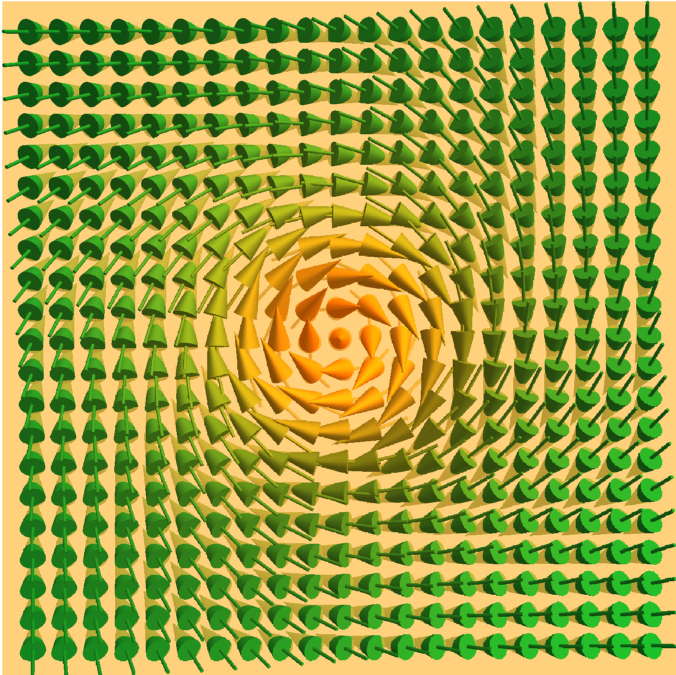}
\caption{Belavin-Polyakov skyrmion with $\gamma = \pi/2$.}
\label{skyrmion} 
\end{figure}

The BP skyrmion field ${\bf s}({\bf r})$ with ${\bf s}^2 = 1$, satisfying the condition ${\bf s} = -\hat{z}$ at infinity, is given by 
\begin{equation}
{\bf s}_{BP}=\left(\frac{2{\lambda} {r}\cos(\phi + \gamma)}{{\lambda}^{2} + {r}^2},\,\frac{2{\lambda} {r}\sin(\phi + \gamma)}{{\lambda}^{2} + {r}^2}\,, \frac{{\lambda}^2-{r}^{2}}{{\lambda}^2+{r}^{2}}\right),
\label{BP}
\end{equation}
where $\lambda$ can be interpreted as the size of the skyrmion and $\gamma$ is a chirality angle. The first image of the N\'{e}el skyrmion ($\gamma = 0$) known to the author appeared in Ref.\ \onlinecite{Kosevich-book}.  The solution with $\gamma = \pi/2$ (Bloch skyrmion) is shown in Fig.\ \ref{skyrmion}. The exchange energy,
\begin{equation}
\mathcal{H}_{ex} =  \frac{J}{2}\int {dxdy} \, \left(\partial_{i}\mathbf{s} \cdot \partial_{i}\mathbf{s}\right),
\label{H-ex}
\end{equation} 
of the BP skyrmion is independent of $\lambda$ and $\gamma$ and equals 
\begin{equation}
E_{ex} = 4\pi J. 
\end{equation}

When the 2D exchange model is studied on a square lattice, the Hamiltonian acquires an additional term \cite{CCG-PRB2012}
\begin{equation}
\mathcal{H}_{lat} = -\frac{J a^2}{24}  \int
dxdy \left(\partial^2_{i}\mathbf{s} \cdot \partial^2_{i}\mathbf{s}\right) 
\label{H-lat}
\end{equation} 
proportional to the square of the lattice spacing $a$. At $\lambda \gg a$ Eq.\ (\ref{BP}) still provides a good approximation for the skyrmion field. The lattice, however, breaks the scale invariance and contributes 
\begin{equation}
E_{lat} = -\frac{2\pi}{3}J \left(\frac{a}{\lambda}\right)^2
\end{equation}
to the energy of the skyrmion, favoring smaller skyrmions over larger skyrmions. 

With the interactions considered so far, the Lagrangian of the skyrmion, up to a constant, is
\begin{equation}
{\cal{L}} = {\cal{L}}_{WZ} - \mathcal{H}_{lat},
\end{equation}
where 
\begin{equation}
{\cal{L}}_{WZ} = \hbar s \int \frac{rdr}{a^2} d\phi  \,\dot{\Phi}(\cos\Theta + 1)
\label{WZ}
\end{equation}
is the Wess-Zumino (geometrical) term \cite{Lectures}. Equations of motion generated by such Lagrangian are equivalent to the Landau-Lifshitz equations of motion for the spin field \cite{Lectures}. Here $s$ is the spin of the unit cell (we chose $s = 1$), $\hbar s/a^2$ is the 2D density of the spin angular momentum,  $\Theta$ and $\Phi$ are spherical coordinates of the spin field, and  $\phi$ and $r$ are polar coordinates in the 2D plane hosting the skyrmion. For the BP skyrmion $\tan\Phi = \tan(\phi + \gamma)$, so that $\Phi = \phi + \gamma$ and  $\dot{\Phi} = \dot{\gamma}$. Eq.\ (\ref{WZ}) then gives
\begin{equation}
{\cal{L}}_{WZ} = 4\pi \hbar \dot{\gamma}\left(\frac{\lambda}{a}\right)^2 \ln\left(\frac{L}{\lambda}\right), 
\end{equation}
\begin{equation}
{\cal{L}} = 4\pi \hbar \dot{\gamma}\left(\frac{\lambda}{a}\right)^2 \ln\left(\frac{L}{\lambda}\right) + \frac{2\pi}{3}J \left(\frac{a}{\lambda}\right)^2,
\label{Lag}
\end{equation}
where $L$ is the cutoff determined by the size of the 2D system.

The Euler-Lagrange equations are
\begin{equation}
\frac{d}{dt} \frac{\partial {\cal{L}}}{\partial \dot{\gamma}} = \frac{\partial {\cal{L}}}{\partial {\gamma}}, \qquad \frac{d}{dt} \frac{\partial {\cal{L}}}{\partial \dot{\lambda}}=\frac{\partial {\cal{L}}}{\partial {\lambda}}. 
\label{EL}
\end{equation}
They give
\begin{equation}
\lambda = {\rm const}, \quad \dot{\gamma} = \frac{J}{6\hbar\ln(L/\lambda\sqrt{e})}\left(\frac{a}{\lambda}\right)^4.
\label{BP-dynamics}
\end{equation}

Thus, within our simple model, the size of the skyrmion does not change even though smaller skyrmions on a lattice have lower energy. So far the only effect of the lattice has been on the chirality angle that now changes linearly with time at a speed inversely proportional to the fourth power of the skyrmion size. 

It is easy to see that this situation does not change even in the presence of the magnetic field, ${\bf H} = -H\hat{z}$ opposite to the total spin of the skyrmion,
\begin{equation}
{\Sigma}_{BP} = \int \frac{dx dy}{a^2} \,(1 + \hat{z} \cdot {\bf s}_{BP})  = 4 \pi  \left(\frac{\lambda}{a}\right)^2\ln\left(\frac{L}{\lambda}\right).
\label{Stot}
\end{equation}
In practice, such a field is needed to provide the stabilizing boundary condition ${\bf s} = -\hat{z}$ at infinity. Similar to the lattice, it favors smaller skyrmions over larger skyrmions. The corresponding Zeeman Hamiltonian is 
\begin{equation}
{\cal {H}}_Z = g\mu_B {\Sigma}_{BP} H = 4 \pi g\mu_B H \left(\frac{\lambda}{a}\right)^2\ln\left(\frac{L}{\lambda}\right),
\end{equation}
with $g$ being the gyromagnetic factor. (In the presence of the field, the integration cutoff $L \gg \lambda$ is determined by the lateral size of the 2D system or by $\delta_H = \sqrt{J/(g\mu_B H}$, whichever is greater \cite{Amel2018}.) The resulting modification of the Lagrangian is 
\begin{equation}
{\cal{L}}_H = 4\pi \hbar \left(\dot{\gamma} - \omega_H \right)\left(\frac{\lambda}{a}\right)^2 \ln \left(\frac{L}{\lambda}\right) + \frac{2\pi}{3}J \left(\frac{a}{\lambda}\right)^2,
\end{equation}
where
\begin{equation}
\omega_H = \frac{g\mu_B H}{\hbar}
\end{equation}
is the Larmor frequency of spin  $s = 1$ in the magnetic field. The solutions of the equations of motion are now
\begin{equation}
\lambda = {\rm const}, \quad \dot{\gamma} = \omega_H + \frac{J}{6\hbar\ln(L/\lambda\sqrt{e})}\left(\frac{a}{\lambda}\right)^4.
\label{BP-H}
\end{equation}

This solution resembles dynamical skyrmions stabilized by the magnetic anisotropy, introduced in Ref.\ \onlinecite{Kovalev1979}. Notice that the second of Eq.\ (\ref{BP-H}) could have been written immediately from Eq.\ (\ref{BP-dynamics}) by observing that  in the problem of spin precession, the application of the magnetic field is equivalent to switching to a coordinate frame rotating with the angular velocity ${\bm \omega_H} = g\mu_B {\bf H}$. Less trivial is the fact that the effect of the discreteness of the atomic lattice is similar to the effect of the field in generating spin precession. 

The impossibility to induce skyrmion collapse by the magnetic field directed along the $z$-axis is related to the rotational invariance with respect to that axis which preserves the $z$-component of the spin angular momentum. It is similar to the well-known impossibility to induce the motion of the domain wall in a uniaxial ferromagnet by the field directed along the anisotropy axis, even though it would have decreased the energy \cite{Lectures}.

\section{Inclusion of dissipation}
\label{dissipation}

We shall now include dissipation in this simple problem. The case of a weak dissipation was studied in Ref.\ \onlinecite{CCG-PRB2012}. Here we consider an arbitrary dissipation strength. The effect of damping on the skyrmion can be described by the dissipation function \cite{Brown}
\begin{equation}
F = \frac{\hbar}{2}\eta \int \frac{dx dy}{a^2} \, \dot{\bf s}^2 ,
\label{F}
\end{equation}
where $\eta$ is the dimensionless Landau-Lifshitz damping parameter. With the help of Eq.\ (\ref{BP}) one obtains for a BP skyrmion
\begin{equation}
{\dot{\bf s}}^2 = \frac{4r^2}{({\lambda}^{2} + {r}^2)^2} \left({\dot{\lambda}}^{2} + \lambda^2 {\dot{\gamma}}^2\right).
\end{equation}
Substitution of this expression into Eq.\ (\ref{F}) and integration over coordinates give
\begin{equation}
F = 4\pi \hbar \eta \left( \frac{{\dot{\lambda}}^2 + \lambda^2 \dot{\gamma}^2}{a^2}\right) \ln\left(\frac{L}{\lambda\sqrt{e}}\right).
\end{equation}

In the presence of dissipation, the Euler-Lagrange equations must be replaced with
\begin{equation}
\frac{d}{dt} \frac{\partial {\cal{L}}}{\partial \dot{\gamma}} = \frac{\partial {\cal{L}}}{\partial {\gamma}} - \frac{\partial F}{\partial \dot{\gamma}}, \qquad \frac{d}{dt} \frac{\partial {\cal{L}}}{\partial \dot{\lambda}} = \frac{\partial {\cal{L}}}{\partial {\lambda}} - \frac{\partial F}{\partial \dot{\lambda}}.
\end{equation}
Since ${\cal{L}}$ does not depend on $\gamma$ and $\dot{\lambda}$, the above equations reduce to
\begin{equation}
\frac{d}{dt} \frac{\partial {\cal{L}}}{\partial \dot{\gamma}} =  - \frac{\partial F}{\partial \dot{\gamma}}, \qquad \frac{\partial {\cal{L}}}{\partial {\lambda}} = \frac{\partial F}{\partial \dot{\lambda}}\,.
\end{equation}
For the smallest skyrmions, whose collapse is dominated by the discreteness of the atomic lattice, they give
\begin{equation}
\frac{d\lambda}{dt} = -\eta \lambda \dot{\gamma}, \qquad  \dot{\gamma} - \eta \frac{\dot{\lambda}}{\lambda} = \frac{Ja^4}{6\hbar\lambda^4\ln(L/(\lambda \sqrt{e})},
\label{lambda-dot-gen}
\end{equation}
which results in 
\begin{equation}
\dot{\lambda} = -\frac{1}{6}\left(\frac{\eta}{1+ \eta^2}\right)\frac{J a^4}{\hbar \lambda^3 \ln(L/(\lambda \sqrt{e})},
\end{equation}
\begin{equation}
\left(\frac{\lambda}{a}\right)^4 \ln\left(\frac{L}{\lambda e^{1/4}}\right)  = \left(\frac{\lambda_0}{a}\right)^4 \ln\left(\frac{L}{\lambda_0 e^{1/4}}\right) - \frac{2\eta J t}{3\hbar (1+ \eta^2)},
\label{small-lambda}
\end{equation}
where $\lambda_0$ is the initial skyrmion size. 

According to Eq.\ (\ref{small-lambda}) the lifetime of the skyrmion is proportional to the fourth power of the initial skyrmion size,
\begin{equation}
t_c = \frac{3\hbar(1+ \eta^2)}{2\eta J }\left(\frac{\lambda_0}{a}\right)^4 \ln\left(\frac{L}{\lambda_0 e^{1/4}}\right).
\label{tc}
\end{equation}
It has a minimum on the damping parameter $\eta$ that occurs at $\eta = 1$. This is easy to understand by noticing that high damping slows down skyrmion collapse by imposing a drag on its motion, while the collapse of a skyrmion at weak damping is inhibited by the conservation of the spin angular momentum.  

An interesting observation is that damping does not slow down the rotation of the chirality angle in the process of skyrmion collapse, 
\begin{equation}
\dot{\gamma} =  \frac{1}{6(1+ \eta^2)}\frac{J a^4}{\hbar [\lambda(t)]^4\ln\{L/[\lambda(t) \sqrt{e}]\}}. 
\end{equation}
On the contrary, with or without damping, it accelerates as the skyrmion size decreases. 

The frequency $J/\hbar$ is typically of order $10^{13}$s$^{-1}$, which gives $t_c \sim 10^{-8}$s for a nanometer-size skyrmion at $\eta \sim 0.1$. This suggests that the collapse of a skyrmion in a pure exchange model is rather slow. It would be a problem if deleting a skyrmion becomes a part of information processing.

\section{Inclusion of DMI}
\label{DMI}

To break with the slow dynamics of skyrmions exhibited by the pure exchange model we should now introduce DMI that breaks the inversion symmetry together with the rotational symmetry, 
\begin{equation}
{\cal{H}}_{DMI} = A\int \frac{dxdy}{a} \left[\left({\bf s} \times \partial_x {\bf s}\right)\cdot \hat{x} + \left({\bf s} \times \partial_y {\bf s}\right)\cdot \hat{y}\right].
\label{Eq-DMI}
\end{equation}
Here $A$ is a parameter having dimensionality of energy that describes the strength of the DMI. For certainty, we chose it to be positive. Since DMI is formed by the combined effect of the exchange and spin-orbit interactions, $A$ is typically small compared to the exchange energy $J$ but large compared to the Zeeman energy. It was tested numerically \cite{Amel2018} that the change in the skyrmion shape due to the DMI is insignificant, especially for small skyrmions studied here. Substituting Eq.\ (\ref{BP}) into Eq.\ (\ref{Eq-DMI}), one obtains 
\begin{equation}
{\cal{H}}_{DMI} = -4\pi A \left(\frac{\lambda}{a} \right) \sin \gamma
\end{equation} 
for the DMI energy of the BP skyrmion. 

\begin{figure}[h]
\centering{}\includegraphics[width=12cm]{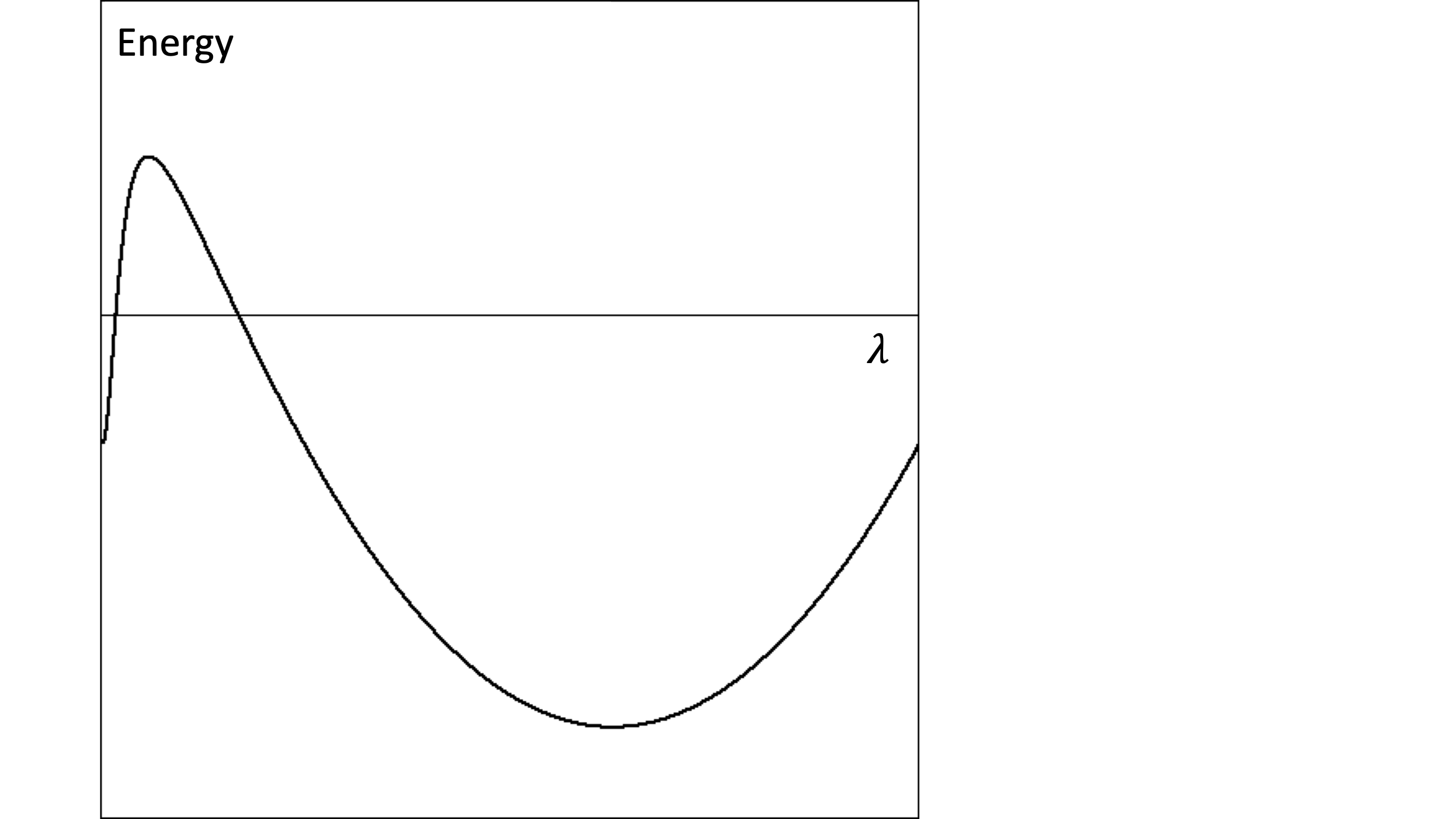}\
\caption{Schematic representation of the skyrmion energy vs size in a model with DMI and magnetic field.}
\label{energy} 
\end{figure}
With the account of contributions from the atomic lattice, DMI, and the magnetic field, the total energy of the skyrmion, up to a constant $4\pi J$ provided by the exchange,  is
\begin{eqnarray}
{\cal{H}} & = & - \frac{2\pi}{3}J \left(\frac{a}{\lambda}\right)^2 - 4\pi A \left(\frac{\lambda}{a} \right) \sin \gamma \nonumber \\
& + & 4 \pi g\mu_B H \left(\frac{\lambda}{a}\right)^2\ln\left(\frac{L}{\lambda}\right) .
\label{total-E}
\end{eqnarray}
It favors a Bloch skyrmion with $\gamma = \pi/2$  depicted in Fig.\ \ref{skyrmion}. The dependence of the energy on the skyrmion size is shown schematically in Fig.\ \ref{energy}. Notice that only $\lambda > a$ has physical meaning. 

To estimate the spatial and time scales involved, it is convenient to use the reduced form of the energy
\begin{equation}
\bar{\cal{H}} = \frac{\cal{H}}{4\pi A} = - \frac{\kappa}{2\bar{\lambda}^2} - \bar{\lambda} \sin \gamma + \frac{1}{2}\delta \bar{\lambda}^2,
\end{equation}
where we replaced the logarithm with a constant $l$ and introduced dimensionless
\begin{equation}
\kappa = \frac{J}{3A} \gg 1, \qquad \delta(H) = \frac{2g\mu_B l H}{A} \ll 1, \qquad \bar{\lambda} = \frac{\lambda}{a}.
\end{equation}
The energy has a minimum at $H < H_c$ which corresponds to $\delta < \delta_c = 2^{1/3} 3/(8 l \kappa^{1/3})$.  As $H$ decreases, the equilibrium skyrmion size at the minimum goes up from ${\bar{\lambda}}_c  = (4\kappa)^{1/3}$ at the field threshold $H = H_c$ to ${\bar\lambda}_H = 1/\delta(H)$ at $H < H_c$.

The reduced Lagrangian is
\begin{equation}
\bar{\cal{L}} = \frac{\cal{L}}{4\pi A} = \frac{1}{2} \dot{\gamma}_{\tau} \bar{\lambda}^2  + \frac{\kappa}{2\bar{\lambda}^2} + \bar{\lambda} \sin \gamma - \frac{1}{2}\delta \bar{\lambda}^2 
\label{Lag+A reduced}
\end{equation}
where
\begin{equation}
\dot{\gamma}_{\tau} = \frac{d\gamma}{d\tau}, \qquad \tau = \frac{A}{2\hbar l}t.
\end{equation}

Skyrmions nucleated to the left of the energy maximum in Fig. \ref{energy} collapse. If the skyrmion of size $\bar{\lambda}_0 < \bar{\lambda}_H$ nucleates to the right from the energy maximum, it begins to expand toward the energy minimum corresponding to $\bar{\lambda} = \bar{\lambda}_H$. We are interested in the characteristic time of such an expansion. Noticing that it is dominated by the DMI, we can start by considering the dynamics of a small skyrmion due to the DMI alone, described by the Lagrangian
\begin{equation}
\bar{\cal{L}} = \frac{1}{2} \dot{\gamma}_{\tau} \bar{\lambda}^2  + \bar{\lambda} \sin \gamma .
\label{Lag-DMI}
\end{equation}
The corresponding equations of motion are
\begin{equation}
\dot{\bar{\lambda}}_{\tau} =  \cos \gamma, \qquad \dot{\gamma}_{\tau} \bar{\lambda}  =  - \sin \gamma.
\label{arb}
\end{equation}
They have an exact solution 
\begin{equation}
\bar{\lambda} = \sqrt{\bar{\lambda}_0^2 +  \tau^2}, \qquad \tan\gamma = \frac{\bar{\lambda}_0}{\tau}.
\label{expansion}
\end{equation}
As the skyrmion expands from $\bar{\lambda} = \bar{\lambda}_0$ at $t = 0$ to $\bar{\lambda} \gg \bar{\lambda}_0$ at $\tau \gg \bar{\lambda}_0$, the chirality angle changes from $\pi/2$ to $0$. 

According to Eq.\ (\ref{expansion}), the expansion of the skyrmion towards equilibrium size $\bar{\lambda}_H$ requires time $\tau_H \sim \bar{\lambda}_H$. Translated into the real time it corresponds to
\begin{equation}
t \sim \frac{2\hbar l} {A}\tau_H = \frac{2\hbar l} {A\delta(H)} = \frac{1}{\omega_H}.
\end{equation}

Damping can be added to the problem along the lines of the previous section. Equations (\ref{arb}) become
\begin{equation}
\dot{\bar{\lambda}}_{\tau} +\eta \dot{\gamma}_{\tau}\bar{\lambda} =  \cos \gamma, \qquad \dot{\gamma}_{\tau} \bar{\lambda} - \eta \dot{\bar{\lambda}}_{\tau} =  - \sin \gamma.
\label{arb-did}
\end{equation}
They possess the same solution for skyrmion expansion induced by the DMI as before, $\bar{\lambda} = \sqrt{\bar{\lambda}_0^2 + \tau_{\eta}^2}$, but with $\tau_{\eta} = \tau/\sqrt{1+\eta^2}$ instead of $\tau$.  This means that dissipation reduces the speed of the expansion by a factor $\sqrt{1+\eta^2}$, which is insignificant at weak damping. 

\section{Conclusion}
\label{conclusion}

We have studied the collapse of skyrmions in centrosymmetric ferromagnetic films and the expansion of skyrmions in 2D films with broken inversion symmetry. These problems determine the times needed for deleting and creating skyrmions if they are to be used as topologically protected memory units for information processing. In systems with rotational invariance, the collapse of skyrmions is rather slow and is determined entirely by the damping. The fastest collapse occurs at the (Landau-Lifshitz) damping constant $\eta = 1$. 

In systems with DMI, the dynamics is faster. To nucleate a skyrmion, one should overcome an energy barrier $U \sim 4 \pi J$. It can be achieved by using a spin-polarized current \cite{Romming}, local heating \cite{Berruto-PRL2018}, a magnetic dipole \cite{Garanin-JAP2018} or a tip of the magnetic force microscope  \cite{Zhang-APL2018}, as well as by temperature \cite{Zhang2020}.  At high temperatures, nucleation of skyrmions should occur naturally with a probability proportional to $\exp(-U/T)$. 

Very small skyrmions collapse. Below a threshold field, a skyrmion above the critical size, nucleated in a film with broken inversion symmetry, expands until it reaches equilibrium size $\lambda_H$. The expansion of skyrmion driven by DMI is barely affected by damping. It requires time $t \sim 1/\omega_H$. This time is in the picosecond range for the fields in the excess of one tesla, which would be sufficiently short for functional skyrmion-based computer memory. 

\section{Acknowledgements}

This work has been supported by the Grant No. DE-FG02-93ER45487 funded by the U.S. Department of Energy, Office of Science.

\end{document}